\documentstyle[12pt]{article}

\begin{document}
\vskip 1cm
\begin{center}
{\large{\bf MASSIVE GAUGE AXION FIELDS}}\\
\vskip 1cm
{\bf P. J. Arias}\\
{\it Grupo de Campos y Part\'{\i}culas, Departamento de F\'{\i}sica, 
Facultad de Ciencias, Universidad Central de Venezuela,\\ AP 47270, 
Caracas 1041-A, Venezuela.}\\
{\bf A. Khoudeir}\\
{\it Centro de Astrof\'{\i}sica Te\'orica, Departamento de F\'{\i}sica,
Facultad de Ciencias, Universidad de los Andes,\\
M\'erida, 5101,
Venezuela.}\\
\end{center}

\begin{abstract}
A gauge invariant formulation for the massive axion is considered. 
The axion acquires mass through a topological term which couples a 
(pseudo)scalar and a third rank antisymmetric tensor. 
Duality, local and canonical equivalences with the non-gauge 
invariant proposal are established. The supersymmetric version of the
gauge invariant model is constructed.
\end{abstract}

It is well known that different theories can be related by duality 
transformations and their classical and quantum properties are
equivalent. 
In four dimensions a massless (pseudo)scalar field: the axion, is 
dual to the antisymmetric field $B_{mn}$ (only if derivative 
couplings are considered). This fact is a 
particular case of the general duality between $p$ and $D-p-2$ forms 
in $D$ dimensions. In this sense, it is known that non-perturbative
effects
break the Pecci-Quinn (PQ) symmetry of the axion, giving mass to it.
Then, a
natural question arises, concernig about what happens with the
duality symmetry. This item was an enigma until two
independent approaches were developed recently  \cite{linde},
\cite{quev}.
In these references the authors show that, though the PQ symmetry is
broken,
duality symmetry still remains. The $B_{mn}$ field is replaced by a
massive
antisymmetric field of third rank $C_{mnp}$. A characteristic feature
of this duality is the lost of abelian gauge invariance for the new
antisymmetric field. The duality now corresponds to a particular case of
the
usual duality between massive $p$ and $D-p-1$ forms in D dimensions.

The conection between a massive (pseudo)scalar field, $\phi$, and a
massive antisymmetric field, $C_{mnp}$ has been considered before in
different contexts \cite{at1}\cite{aur}\cite{curt}\cite{raje}.
In this letter, we will show that a gauge invariant model, which
involves a topological coupling, considered in reference \cite{aur}, in
the
context of the $U(1)$ problem, is locally equivalent to the non gauge
invariant
proposal for the massive axion. This equivalence is analogous to that
between the ``self-dual'' \cite{djvn} and topologically massive
\cite{tm}
models in three dimensions, in the same manner as the equivalence
between the
Proca and the topologically massive gauge invariant theories in four
dimensions \cite{pio} \cite{tm4}. In the preceding examples, in three
and
four dimensions, the connection occurs between a gauge sensitive and a
gauge
invariant model. It has been shown that these models are locally
equivalent 
and one of them is a gauge fixed version of the other
\cite{rita}\cite{pio}. 
Nevertheless the global equivalence depends on the cohomological
structure of 
the base manifold \cite{pja}\cite{pio}\cite{jorge}\cite{sen}.
For the models that we compare in this letter we will
face this dual equivalence in several ways. One of them, described below
by $I_{M1}$,
is a first order version of the massive (pseudo)scalar model. The other
model,
described below by $I_{M2}$, is gauge invariant.
We will note that the space of solutions of both models differ in a
topological
sector that can be gauged out on trivial manifolds. The presence of this
topological
sector appears explicitly in the partiton function of the gauge
invariant
model as a topological factor related with the cohomological
structure of the base manifold.
We will prove (canonically) that one of the models is a gauge
fixed version of the gauge invariant one. In particular,
we will see the equivalence through the existence of a first order
master
action, that will allow us to consider global aspects and interactions.
Finally we will construct a supersymmetric formulation of the gauge
invariant 
action of the massive axion.

An illustrative model for the massive axion is given by the following 
master action\cite{curt} \cite{quev1}
\begin{equation}
I = <\frac{1}{2}v_m v^m + \phi\partial_{m}v^{m} - \frac{1}{2}m^2 \phi^2
>,
\end{equation}
where $v_m$ is a vector field and $\phi$ is a scalar field ($< >$ 
denotes integration in four dimensions).
Eliminating the field $v_m $ through its equation of motion($v_m = 
\partial_m\phi$), the action for a massive scalar is obtained, while
using 
the equation of motion of the scalar field $\phi$ 
($\phi = \frac{1}{m^2}\partial_m v^m$), we have 
\begin{equation}
I_v = \frac{1}{2}< v_m v^m + \frac{1}{m^2}(\partial_m v^m)^2 >.
\end{equation}
The propagator corresponding to the field $v_m$ is 
$\eta_{mn} - \frac{k_{m}k_{n}}{k^2 + m^2}$, which is just equal to those 
discussed in \cite{linde}. 
If we introduce  the dual of the vector field: 
$v^m = \frac{1}{3!}m\epsilon^{mnpq}C_{npq}$, 
we have the following master action
\begin{equation}
I_{M1} = < -\frac{1}{2.3!}m^2 C^{mnp}C_{mnp} +
\frac{1}{3!}m\epsilon^{mnpq}
\phi\partial_{m}C_{npq} - \frac{1}{2}m^2 \phi^2 >,
\end{equation}
from which duality is easily infered.
To see this, we can use its equations of motion
\begin{equation}
mC^{mnp} - \epsilon^{mnpq}\partial_q \phi = 0,
\label{eqC}
\end{equation}
\begin{equation}
m\phi - \frac{1}{3!}\epsilon^{mnpq}\partial_m C_{npq} = 0,
\label{eqfi}
\end{equation}
to eliminate $C_{mnp}$ (or $\phi$), thus obtaining the action for a
massive scalar field (or the massive antisymmetric field $C_{mnp}$). 
In any case, there is not gauge invariance at all. 
We can ask whether there really exist a gauge invariant 
theory compatible with a massive term for the axion field. 
The answer is positive, and the action is
\begin{equation}
I_{M2} = <-\frac{1}{2}\partial_m\phi\partial^m\phi -
\frac{1}{2.4!}G_{mnpq}
G^{mnpq} - \frac{m}{3!}\epsilon^{mnpq}C_{mnp}\partial_q\phi >,
\end{equation}
where $G_{mnpq} \equiv \partial_m C_{npq} - \partial_n C_{mpq} + 
\partial_p C_{mnq} - \partial_q C_{mnp}$ is the field strength
associated 
with the antisymmetric field $C_{mnp}$. This action was considered
previously 
in reference \cite{aur} as a generalization to four dimensions of the 
Schwinger model in two dimensions. This action is invariant, up
to a surface term, under the abelian gauge transformations
\begin{equation}
\delta_{\xi} C_{mnp} = \partial_{m}\xi_{np} + \partial_{n}\xi_{pm} + 
\partial_{p}\xi_{mn}, \quad \delta_{\xi}\phi = 0.
\end{equation}
and under the global transformation: $\delta\phi = constant$.
In $I_{M_2}$ we note that the coupling term is a BF topological term.
This term preserves the number of degrees of freedom of the
non-interacting terms (just the one
connected with the massless (pseudo)scalar field), but change the
physical
characteristic of the spectrum (the (pseudo)scalar field becomes
massive).
This situation is like the one in three dimension with the topologically
massive model where the topological Chern-Simons term changes the
physical
properties of the Maxwell field (which in three dimensions is a massless
scalar) to a massive spin 1 theory. Other example of this situation, in
four dimensions, occurs in the Cremmer-Scherk\cite{cs} or topologically
massive 
model where the original fields are a massless scalar (one degree of
freedom) 
and a photon with helicity 1 (two degrees of freedom) and the resulting
theory
with the topologiacal interaction term is a massive spin 1 excitation
(three degrees of freedom).

The connection of $I_{M_2}$ and the massive (pseudo)scalar is
straightforward: introducing $\lambda \equiv
-\frac{1}{4}\epsilon^{mnpq}
G_{mnpq}$ as the dual of the strenght field $G_{mnpq}$ into the action
$I_{M2}$, 
we observe that $\lambda$ plays the role of an auxiliary field, whose 
elimination through its equation of motion ($\lambda = -m\phi$) leads to 
the action of a massive (pseudo)scalar field. 

It is worth, at this point, to note that since $I_{M_2}$ depends only of
derivatives of the scalar field, then its dual theory can be achieved,
replacing $\partial_m \phi$ by $\frac{1}{2}l_{m}$ and adding a BF term: 
$\frac{1}{4}l_m\epsilon^{mnpq}\partial_n B_{pq}$\cite{nt}. The dual
action will be \cite{aur}\cite{raje}
\begin{equation}
I_d = < -\frac{1}{2.4!}G_{mnpq}G^{mnpq} - \frac{1}{2.3!}
(mC_{mnp} - H_{mnp})(mC^{mnp} - H^{mnp}) >,
\end{equation}
where $H_{mnp} = \partial_m B_{np} + \partial_n B_{pm} + \partial_p
B_{mn}$ is 
the field strength of the antisymmetric field $B_{mn}$, which was
introduced 
in the BF term. This action describes the interaction of open
membranes 
whose boundaries are closed strings \cite{at} and is invariant under the 
following gauge transformations
\begin{equation}
\delta C_{mnp} = \partial_{m}\xi_{np} + \partial_{n}\xi_{pm} + 
\partial_{p}\xi_{mn}, \quad \delta B_{mn} = \partial_m\lambda_n - 
\partial_n\lambda_m - m\xi_{mn}.
\end{equation}
The $\xi$ gauge transformation allows us gauged away the antisymmetric
field 
$B_{mn}$, leading to the massive antisymmetric field $C_{mnp}$ action.

Let us, now, see that $I_{M_2}$ is locally equivalent to $I_{M1}$
on-shell.
The equations of motion for $I_{M2}$ can be written, suggestively, as
\begin{equation}
\epsilon^{mnpq}\partial_q (mC_{mnp} - \epsilon_{mnpr}\partial^{r}\phi )
= 0, 
\label{eqfiinv}
\end{equation}
\begin{equation}
\epsilon^{mnpq}\partial_q (m\phi - \frac{1}{3!}\epsilon^{rstu}
\partial_{r}C_{stu} ) = 0.
\label{eqCinv}
\end{equation}
 These equations, are just the result of aplying the differential
operator
 $\epsilon^{mnpq}\partial_q$ to the equations of motion derived 
from $I_{M1}$: eqs. (\ref{eqC}) and (\ref{eqfi}).
 From here the local equivalence can be stated: equations
(\ref{eqfiinv}) and
 (\ref{eqCinv}) can be written locally as
\begin{equation}
mC_{mnp} - \epsilon_{mnpr}\partial^{r}\phi = \partial_m \lambda_{np}
+ \partial_n \lambda_{pm} + \partial_p \lambda_{mn},
\label{deltaC}
\end{equation}
and 
\begin{equation}
m\phi - \frac{1}{3!}\epsilon^{mnpq}\partial_{m}C_{npq} = constant.
\label{deltafi}
\end{equation}
Now, the right sides of (\ref{deltaC}) and (\ref{deltafi}) correspond,
respectively, to gauge changes in $C_{mnp}$ and $\phi$. So those sides
are
gauge equivalent to zero, leaving us with equations (\ref{eqC}) and
(\ref{eqfi}). In other words the physical relevant solutions of both
models are
the same, at the local level.
Equivalently, closed 0 and 3-forms $(\partial_m\phi = 0 =
\partial_{[m} C_{npq]} )$ are always solutions of the 
equations (\ref{eqfiinv}) and (\ref{eqCinv}), while they
become trivial in the model described by $I_{M1}$. 
Locally, this closed forms are always exact and then the gauge
invariance allow us to gauge them away. Globally, this will depend on
the 
cohomological structure of the base manifolds and both models may differ 
in their space of solutions. We will see later that this topological
dependence
will appear explicitly in the partition function.

Proceeding with the comparison we will show the canonical equivalence
for these two theories on cohomological trivial region of space-time.
For the non-gauge
invariant 
action $I_{M1}$, the canonical hamiltonian density is found to be 
\begin{equation}
{\cal H}_1 = \frac{1}{2.3!}m^2 C_{ijk}C_{ijk} + \frac{1}{2}m^2 \phi^{2}
+ 
\frac{1}{2}\partial_{i}\phi \partial_{i}\phi ,
\end{equation}
subject to the set of second class constraints
\begin{equation}
\psi = 0 \quad and \quad \psi_{ijk} = \pi_{ijk} -
\frac{1}{3!}m\epsilon_{ijk}\phi, 
\end{equation}
where $\pi$ and $\pi_{ijk}$ are the canonical conjugated momenta
associated to 
$\phi$ and $C_{ijk}$, respectively. 

On the other hand, for the gauge invariant action $I_{M2}$, we have
\begin{equation}
{\cal H}_2 = 3\pi_{ijk}\pi_{ijk} +
\frac{1}{2}\partial_i\phi\partial_i\phi 
+ \frac{1}{2}[\pi - \frac{m}{3!}\epsilon^{ijk}C_{ijk}]^{2},
\end{equation}
subject to the first class constraints
\begin{equation}
\theta_{jk} = \partial_i\pi_{ijk} -
\frac{1}{3!}m\epsilon^{ijk}\partial_i\phi .
\end{equation}
These constraints are doubly reducible, i.e., there are functions 
$Z_{k}^{ij}(x,y)$ and $Z_{k}(x,y)$ such that 
$\int dy Z_{k}^{ij}(x,y)\theta_{ij}(y) = 0$ and 
$\int dy Z^{k}(x,y)Z_{k}^{ij}(x,z) = 0$. This fact must be taken into 
account when we want the gauge freedom to be fixed.

It is straightforward to show that these two hamiltonian densities are
related by linear combinations of the set of second class constraints of
${\cal H}_1$, 
reflecting the canonical equivalence:
\begin{equation}
{\cal H}_2 = {\cal H}_1 + 3\psi_{ijk}[\psi_{ijk} +
\frac{1}{3}m\epsilon_{ijk}\phi ] + 
\frac{1}{2}\psi [\psi - \frac{1}{3}m\epsilon_{ijk}C_{ijk}].
\label{H1H2}
\end{equation}
This situation is similar to those theories which involve a mass
mechanism 
through topological coupling \cite{rita}\cite{pio}\cite{pja}. Moreover,
we can 
see that $I_{M1}$ is a gauge fixed version of the gauge invariant 
action $I_{M2}$. In fact, we note that $\theta_{ij} =
\partial_k\psi_{ijk}$ 
can be considered as the first class constraints and $\psi = 0$ as the 
gauge fixing condition$\footnote{$\psi_{ijk}$ and $\partial_i\psi_{ijk}$ 
have the same information in three space dimensions}$. In the latter
case
the first order hamiltonian results to be $H_2$. The canonical procedure
continues at this point and then the quantum equivalence is clear.

Now, in order to touch further aspects it is worth looking at the
equivalence
through the following
master action
\begin{eqnarray}
I_M = &<&-\frac{1}{2.3!}m^2 a_{mnp}a^{mnp} - \frac{1}{2!}m^2\psi ^2 + 
\frac{1}{4!}m\epsilon^{mnpq}\psi G_{mnpq} \nonumber \\ 
&+& \frac{1}{3!}m\epsilon^{mnpq}(a_{mnp} - C_{mnp})\partial_q\phi>.
\end{eqnarray}
Independent variations in $a_{mnp}, \psi, C_{mnp}$ and $\phi$ lead to
the 
following equations of motion
\begin{equation}
a^{mnp} = \frac{1}{m}\epsilon^{mnpq}\partial_q\phi,
\end{equation}
\begin{equation}
\psi = \frac{1}{4!m}\epsilon^{mnpq}G_{mnpq},
\end{equation}
\begin{equation}
\epsilon^{mnpq}\partial_{m}(\psi - \phi) = 0
\end{equation}
and 
\begin{equation}
\epsilon^{mnpq}\partial_{q}(a_{mnp} - C_{mnp}) = 0.
\end{equation}
Replacing the expressions for $a_{mnp}$ and $\psi$ given by eqs. (20)
and (21) 
into $I_M$, the gauge invariant action $I_{M2}$ is obtained. 
On the other hand, the solutions of the equations of motion 
(22) and (23) are
\begin{equation}
\phi - \psi = \omega, \quad C_{mnp} - a_{mnp} = \Omega_{mnp},
\end{equation}
where $\omega$ and $\Omega_{mnp}$ are $0$ and $3$-closed forms,
respectively. 
Locally, we can set 
\begin{equation}
\omega = constant, \quad \Omega_{mnp} \equiv L_{mnp} = \partial_m l_{np}
+ 
\partial_n l_{pm} + \partial_p l_{mn},
\end{equation}
and substituting into $I_M$, we obtain the following
``St\"{u}eckelberg''
type action
\begin{eqnarray}
I_s = &<&-\frac{1}{2.3!}m^2 (C_{mnp} - L_{mnp})(C^{mnp} - L^{mnp}) -
\frac{1}{2}
m^2 (\phi - \omega)^2 \nonumber \\ 
&+& \frac{1}{4!}m\epsilon^{mnpq}(\phi - \omega)G_{mnpq}>.
\end{eqnarray}
This action is invariant under 
\begin{equation}
\delta_{\xi} C_{mnp} = \partial_{m}\xi_{np} + \partial_{n}\xi_{pm} + 
\partial_{p}\xi_{mn}, \quad \delta_{\xi}l_{mn} = \xi_{mn},
\end{equation}
which allow us to gauge away the $l_{mn}$ field and recover $I_{M1}$ (we 
have redefined $\phi -\omega$ as $\phi$ since $\omega$ is a constant).
In 
this way, the local equivalence is also stated. On the other hand, we
can 
consider $\psi = \phi - \omega$ and $a_{mnp} = C_{mnp} - \Omega_{mnp}$ 
as general solutions in order to obtain the following gauge
invariant 
action
\begin{eqnarray}
{\bar I}_{M} = &<&-\frac{1}{2.3!}m^2 (C_{mnp} - \Omega_{mnp})(C^{mnp} - 
\Omega^{mnp}) - 
\frac{1}{2!}m^2(\phi - \omega) ^2 \nonumber \\ 
&+& \frac{1}{4!}m\epsilon^{mnpq}(\phi - \omega) G_{mnpq} - 
\frac{1}{3!}m\epsilon^{mnpq}\Omega_{mnp}\partial_q\phi>.
\label{II}
\end{eqnarray}
This action is global and locally equivalent to $I_{M2}$, so
the topological sectors, not present in $I_{M1}$, are now included.
Indeed, the equations of motion which are obtained after performing 
independent variations on $C_{mnp}, \Omega_{mnp}, \phi$ and $\omega$ in 
${\bar I}_{M}$ are
\begin{eqnarray}
m(C^{mnp} - \Omega^{mnp}) &-& \epsilon^{mnpq}\partial_q (\phi - \omega ) 
= 0 \nonumber \\ 
m(C^{mnp} - \Omega^{mnp}) &-& \epsilon^{mnpq}\partial_q \phi = 0
\end{eqnarray}
and 
\begin{eqnarray}
m(\phi - \omega) &-& \frac{1}{3!}\epsilon^{mnpq}\partial_m C_{npq} 
+ \frac{1}{3!}\epsilon^{mnpq}\partial_m\Omega_{npq} = 0 \nonumber \\ 
m(\phi - \omega) &-& \frac{1}{3!}\epsilon^{mnpq}\partial_m C_{npq} = 0,
\end{eqnarray}
from which is easily deduced that $\omega$ and $\Omega_{mnp}$ are 
closed 0 and 3-forms, respectively, i.e.
$\epsilon^{mnpq}\partial_q\omega 
= 0 = \epsilon^{mnpq}\partial_m\Omega_{npq}$. Taking into account this
last 
result and applying the differential operator $\epsilon^{mnpq}\partial_q
$ on 
the set first order differential equations given by (30), the equations
of 
motion for the gauge invariant action $I_{M2}$ are obtained (equations
(\ref{eqfiinv}) and
(\ref{eqCinv})). We can say that ${\bar I}_{M}$ is the correct
modification
to $I_{M_1}$ in
order to include the, originally missing, topological sectors. This
aspect will
be important when interactions are to be considered.
Moreover, we can eliminate $\phi$ and $C_{mnp}$ to achieve
\begin{equation}
{\bar I}_{M1} = I_{M1[a,\psi]} - I_{top[\omega ,\Omega]},
\end{equation}
where
\begin{equation}
I_{top[\omega ,\Omega]} = < \frac{1}{3!}m\epsilon^{mnpq}\Omega_{mnp}
\partial_{q}\omega >
\label{part}
\end{equation}
is the BF term for the topological coupling between 
$0$ and $3$-forms in four dimensions.
In other direction, if we couple a source to
$\phi$ and $C_{mnp}$ only, in (\ref{II}), it is easy to see that the sum
of
any solution of the inhomogeneous equations of $I_{M_1}$ with any
solution of the inhomogeneous equation of the BF theory is a solution of
the inhomogeneous equation of $I_{M_2}$.

From (\ref{part}) we have that
the partition funtions of $I_{M1}$ and $I_{M2}$ differ by a topological
factor.
\begin{equation}
Z_{M2} = Z_{top}Z_{M1}
\end{equation}
This topological factor is associated with the topologiacal sectors
not present in the space of solutions of $I_{M1}$. In general, on
manifolds with non trivial topological structure
$Z_{top} \neq 1$. Only when the manifold has a trivial structure, 
we will have $Z_{top} \equiv 1$, reflecting the local and global
equivalence
(In the cases of duality equivalence taking into account global 
aspects see references \cite{allen} \cite{rest}).

Finally, we will construct the supersymmetric formulation 
of the gauge invariant action for the massive axion. For this 
proposal, we need (anti)chiral superfields ($\Phi$ and $\Phi^{+}$) 
and a superfield $V$ whose supersymmetric field strength is 
$G = -\frac{1}{2}(DD + {\bar D}{\bar D})V$, where ($DDV$) 
${\bar D}{\bar D}$ is (anti) chiral superfield. This later superfield 
describes the $C_{mnp}$ in a supersymmetric way\cite{gates}. 
The supersymmetric action is 
\begin{eqnarray}
I &=& \frac{1}{16}\int d^4 x d^2\theta d^2 {\bar \theta} G^2 + 
\frac{1}{2}\int d^4 x d^2\theta d^2 {\bar \theta} \Phi \Phi^{+}
\nonumber\\
&+& \frac{1}{4}m\int d^4 x d^2\theta \Phi {\bar D}{\bar D}V 
+ \frac{1}{4}m\int d^4 x d^2 {\bar \theta} \Phi^{+}DDV .
\end{eqnarray}
where the last two terms are the supersymmetric extension of the 
topological coupling. In components (using Majorana spinors), this 
action is written down as
\begin{eqnarray}
I &=& -\frac{1}{2}\partial_m A \partial_m A - \frac{1}{2}\partial_m B 
\partial_m B + \frac{1}{2}(\partial_m v_m )^2 + \frac{1}{2}d^2 
-\frac{i}{2}{\bar \lambda}\gamma^{m}\partial_m\lambda \nonumber\\ 
&-&\frac{1}{2}\partial_m a \partial_m a - \frac{1}{2}\partial_m \phi 
\partial_m \phi -\frac{i}{2}{\bar \chi}\gamma^{m}\partial_m\chi 
+ \frac{1}{2}f^2 + \frac{1}{2}g^2 \\ \nonumber
&-&m[\phi\partial_m v_m + ad + i{\bar \chi}\gamma_{5}\lambda - 
(fB - gA)],
\end{eqnarray}
where $(A,B,v^m \equiv \frac{1}{3!}\epsilon^{mnpq}C_{npq},d,\lambda )$ 
and $(a,\phi ,\chi ,f,g)$ are the components of V and $\Phi$ multiplets, 
respectively. This action describes 8 bosonic and 8 fermionic degrees of 
freedom and is invariant under the off-shell supersymmetric
transformations
\begin{eqnarray}
\delta A &=& i{\bar \epsilon}\lambda , \quad \delta B = i{\bar \epsilon}
\gamma_5 \lambda , \nonumber\\ 
\delta v &=& i{\bar \epsilon}\gamma^m\lambda , \quad 
\delta d= i{\bar \epsilon}\gamma_{5}\gamma^m\partial_m \lambda ,
\nonumber\\
\delta\lambda &=& \epsilon(\partial .v) + \gamma_5\epsilon d + 
\gamma^m \epsilon\partial_m A - \gamma_5\gamma^m \epsilon\partial_m A
\end{eqnarray}
and
\begin{eqnarray}
\delta a &=& i{\bar \epsilon}\chi , \quad \delta\phi = i{\bar \epsilon}
\gamma_5\chi ,\nonumber\\ 
\delta\chi &=& [\partial_m a - \gamma_5\partial_m\phi]\gamma^m\epsilon 
+ [f + \gamma_5 g]\epsilon , \nonumber \\ 
\delta f &=& i{\bar \epsilon}\gamma^m\partial_m\chi , \quad 
\delta g = i{\bar \epsilon}\gamma_5\gamma^m\partial_m\chi .
\end{eqnarray}
Note that the $a,A$ and $B$ bosonic fields also have the same mass as 
the axion field, as can be seen after eliminating the 
$g,d$ and $f$ auxiliary fields, just as the dilaton acquire mass in the
linear
multiplet, in the supersymmetric formulation for
the Cremmer-Sherk action\cite{siegel}

Sumarizing, we have seen that a gauge invariant description for 
massive axions is possible. This gauge invariant description
 is (locally)equivalent to the
non-gauge invariant proposal. 
The supersymmetric formulation was given. 
A complete BRST analysis of the gauge invariant model considered 
in this paper is under consideration.

\begin{center}
{ACKNOWLEDGEMENTS}
\end{center}

One of the authors (AK) would like to thank to the Consejo de Desarrollo 
Cient\'{\i}fico y Human\'{\i}stico de la Universidad de los
Andes(CDCHT-ULA) by institutional support under project C-862-97.

\end{document}